\documentclass[%
 reprint,
 amsmath,amssymb,
 aps
]{revtex4-2}

\usepackage{graphicx}
\usepackage{dcolumn}
\usepackage{bm}
\usepackage{xcolor}
\usepackage{braket}

\begin{document}

\preprint{APS/123-QED}

\title{Experimental demonstration of conjugate-Franson interferometry}

\author{Changchen Chen}
\affiliation{Research Laboratory of Electronics, Massachusetts Institute of Technology, Cambridge, MA 02139 ,USA}

\author{Jeffrey H. Shapiro}
\affiliation{Research Laboratory of Electronics, Massachusetts Institute of Technology, Cambridge, MA 02139 ,USA}

\author{Franco N.C. Wong}
\affiliation{Research Laboratory of Electronics, Massachusetts Institute of Technology, Cambridge, MA 02139 ,USA}

\date{\today}

\begin{abstract}
Franson interferometry is a well-known quantum measurement technique for probing photon-pair frequency correlations that is often used to certify time-energy entanglement. We demonstrate the complementary technique in the time basis, called conjugate-Franson interferometry, that measures photon-pair arrival-time correlations, thus providing a valuable addition to the quantum toolbox. We obtain a conjugate-Franson interference visibility of $96\pm 1$\% without background subtraction for entangled photon pairs generated by spontaneous parametric down-conversion. Our measured result surpasses the quantum-classical threshold by 25 standard deviations and validates the conjugate-Franson interferometer (CFI) as an alternative method for certifying  time-energy entanglement. Moreover, the CFI visibility is a function of the biphoton's joint temporal intensity and is therefore sensitive to that state's spectral phase variation, something which is not the case for Franson interferometry or Hong-Ou-Mandel interferometry. We highlight the CFI's utility by measuring its visibilities for two different biphoton states, one without and the other with spectral phase variation, and observing a 21\% reduction in the CFI visibility for the latter. The CFI is potentially useful for applications in areas of photonic entanglement, quantum communications, and quantum networking.  
\end{abstract}

\maketitle

Time-energy entanglement is the quintessential quantum resource for enabling next-generation quantum technologies such as one-way quantum computation \cite{nielsen2004optical}, quantum-enhanced sensing \cite{zhang2015entanglement, lloyd2008enhanced, zhang2020multidimensional}, and quantum-secured communications \cite{zhong2015photon, lee2014entanglement}. Franson interferometry is a well-known technique for measuring the nonlocal timing coincidence of photon pairs~\cite{franson1989bell}. Because Franson interference visibility resembles the Clauser-Horne-Shimony-Holt (CHSH) inequality, it is often used to characterize the quality of a biphoton's time-energy entanglement \cite{clauser1969proposed}. Nevertheless, Franson interferometry only quantifies the photon pair's correlation in the frequency domain and does not provide correlation information in the time domain \cite{zhang2014unconditional}. Without time-domain characterization, Franson interferometry by itself cannot reveal a full picture of the biphoton's nonclassical correlations. Characterization of entangled photon pairs in the time domain is challenging because there is no readily available experimental method to directly measure two-photon timing correlation. One can extract two-photon time correlation from their joint temporal intensity (JTI) measurements but they typically require sub-picosecond temporal gating and single-photon nonlinear conversion that tend to limit measurement efficiencies~\cite{kuzucu2008joint, davis2018measuring}. 

The conjugate-Franson interferometer (CFI) was proposed as a quantum measurement technique for probing two-photon correlation in the time domain in contrast to the Franson interferometer's frequency-domain probing~\cite{zhang2014unconditional}. The two interferometric techniques form a complementary quantum-measurement duo for quantifying biphotons' time-energy entanglement. Indeed, the CFI was proposed to work together with the Franson interferometer to provide a tighter bound on an eavesdropper's accessible information in high-dimensional quantum key distribution than is achievable with the Franson interferometer alone \cite{zhang2014unconditional}. 

The addition of the CFI to the expanding quantum toolbox offers new or improved measurement capability in quantum photonic studies. Although biphoton spectral phase information can be obtained using frequency-resolved \cite{tischler2015measurement} or time-resolved \cite{chen2015measuring} two-photon local interference, these techniques require nearly-degenerate photon pairs.  The CFI, however, is a nonlocal two-photon measurement that is suitable for nondegenerate photon pairs. Other means to probe temporal correlations include the use of an electro-optic spectral shearing interferometer \cite{davis2020measuring} with femtosecond pulse gating, and phase-sensitive detection with a stable and well-characterized classical field \cite{beduini2014interferometric}. The CFI, on the other hand, does not require a reference field and can work with photon pairs generated by pulsed or continuous-wave (cw) pumping. 

Recent studies on quantum frequency combs have underscored the inability of Hong-Ou-Mandel interference (HOMI) \cite{HOM1987} or Franson interference to distinguish two frequency combs that differ only in their spectral phase content. Lingaraju {\em et al.} made HOMI measurements on biphoton frequency combs with different spectral phase variations and found identical HOMI signatures \cite{lingaraju2019quantum}. To understand what properties of biphoton frequency combs can be extracted by different interferometric measurements, Chang {\em et al.} argues that both HOMI and Franson interference are functions of the biphoton's joint spectral intensity (JSI), whereas the CFI measures the state's JTI \cite{chang2021648}. Because spectral phase variation does not affect the JSI, it confirms the observation in \cite{lingaraju2019quantum} and suggests that the CFI is the appropriate measurement tool to distinguish combs with spectral phase variations. A classical-optics analog is how linear dispersion of a transform-limited optical pulse imposes a phase chirp that results in pulse broadening which is detectable by time-domain but not frequency-domain measurements.  

\begin{figure*}
    \centering
    \includegraphics[scale=0.8]{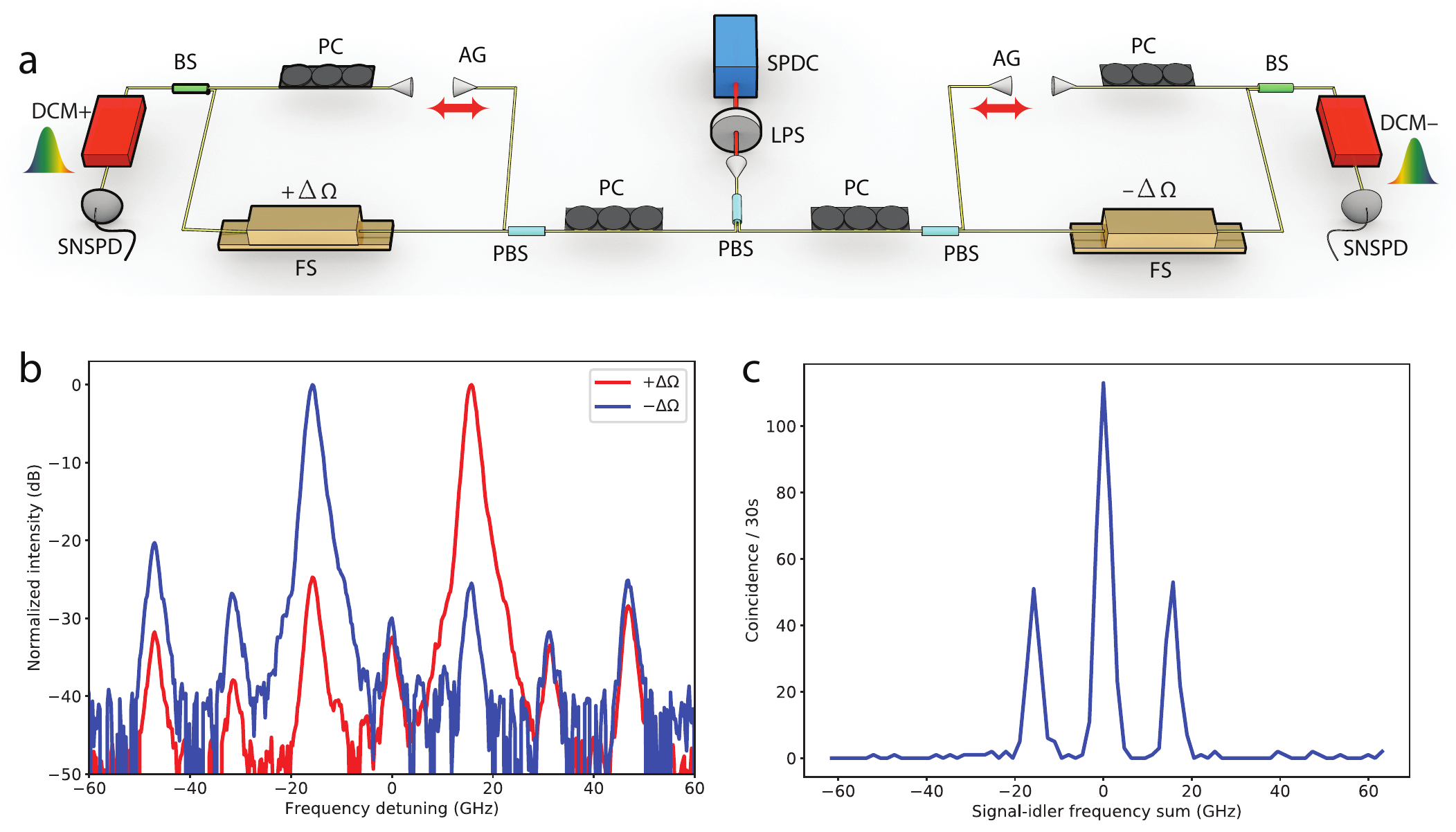}
    \caption{\textbf{a}, Experimental setup of our conjugate-Franson interferometer. Time-energy entangled signal-idler photon pairs generated by cw pumped SPDC were coupled into an optical fiber and routed to their respective MZIs. The fiber-based CFI was placed inside a custom-built two-stage thermal box for phase stabilization. The MZI outputs were detected with SNSPDs and their arrival times recorded for coincidence measurements. LPS: long-pass filter; PBS: polarizing beam splitter; PC: polarization controller; FS: frequency shifter for $\Delta \Omega$ ($-\Delta\Omega$) frequency shift; AG: tunable air gap; BS: 50/50 beam splitter; DCM+($-$): dispersion module with normal (anomalous) dispersion. \textbf{b}, Log-scale display of the frequency shifters' output spectra, measured using classical light, that show signal-to-noise ratios of at least 20\,dB limited by higher-order sidebands. Maximum intensities of both spectra normalized to 0\,dB. \textbf{c}, Measured CFI coincidences: central peak location determines zero detuning of the signal-idler sum frequency. 30-s integration time for each data point; measurement taken with MZI phase sum $\phi_T \approx \pi/2$.}
    \label{fig1}
\end{figure*}

In this Letter, we report implementing the CFI and obtaining a $96\pm1$\% CFI fringe visibility without background subtraction for time-energy entangled photon pairs generated by cw pumped spontaneous parametric down-conversion (SPDC) Our measured visibility surpasses the quantum-classical threshold by $\sim$25 standard deviations, thus validating the CFI as a valuable tool for quantifying a biphoton's time-energy entanglement. Moreover, we demonstrate the CFI's unique capability by utilizing it to distinguish between two  biphoton states that differ only in their spectral phase content, one having a uniform phase and the other with a nonuniform phase. Our CFI measurements show a visibility degradation of 21.2\% for the biphoton state with a nonuniform spectral phase when compared to the visibility obtained with a uniform phase (which is transform limited), in agreement with our theoretical calculation. The visibility degradation indicates a decrease in timing correlation as the result of the presence of spectral phase, whose information cannot be obtained using standard tools for analyzing the joint properties of photon pairs, such as HOMI, Franson interference, and JSI measurements \cite{lingaraju2019quantum, chen2017efficient, zielnicki2018joint}. We expect that the addition of the CFI to the quantum toolbox provides a simpler way to characterize time-domain correlation and a new method to monitor spectral phase information of time-energy entangled photon pairs. Hence we believe the CFI will enhance future developments of entanglement systems for computing, communication, and sensing applications.

\begin{figure*}
    \centering
    \includegraphics[scale=0.8]{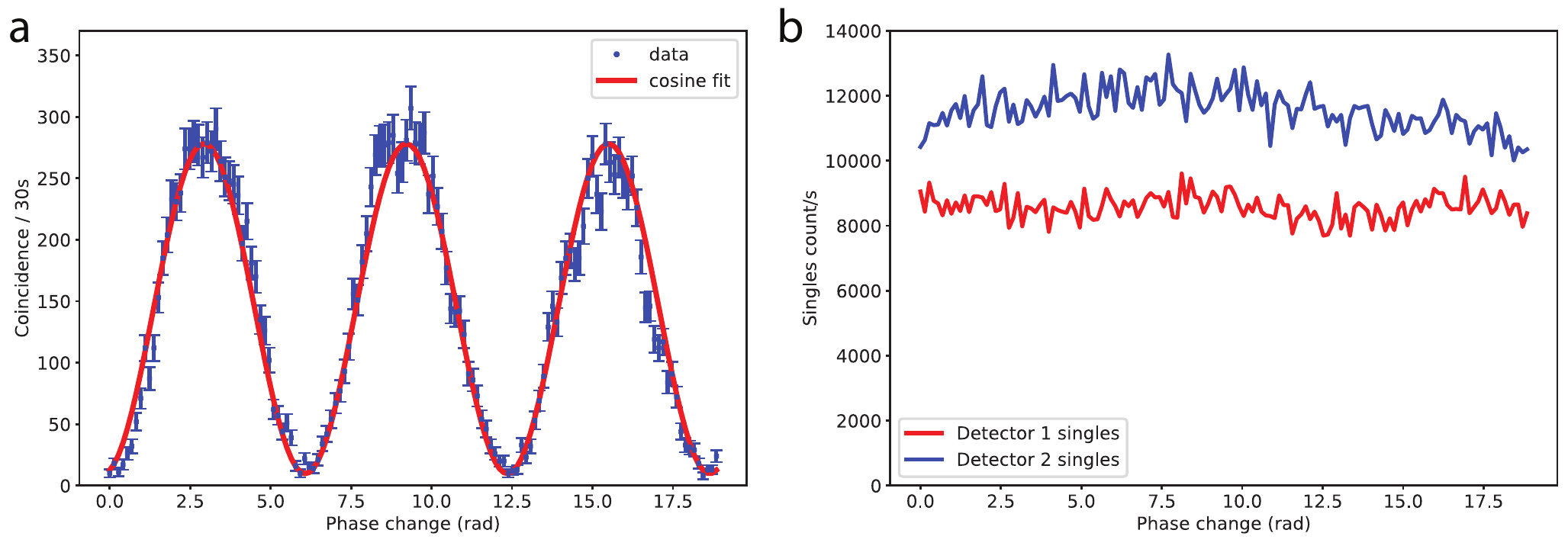}
    \caption{\textbf{a}, Coincidences (blue) as a function of MZI phase sum $\phi_T$, with calculated uncertainties assuming Poisson statistics. Least-squares fit (red) to the form $A[1+V\text{cos}(\phi_T)]$ yields a fitted CFI visibility ($V$) of 93\%. \textbf{b}, Singles count rates for both detectors as functions of the MZI phase sum $\phi_T$, showing no meaningful variations. No background counts are subtracted from measured data.}
    \label{fig2}
\end{figure*}

The conjugate-Franson interferometer comprises two Mach-Zehnder interferometers (MZIs) that are separated in space with each MZI having equal-length arms. For time-energy entanglement characterization, signal (idler) photons of entangled signal-idler photon pairs are sent to one (the other) MZI, and their coincidence outputs are monitored to measure the conjugate-Franson interference. An optical frequency shifter is placed in one of the arms within each interferometer, implementing a $\Delta \Omega$ frequency shift for the signal photons and a $-\Delta \Omega$ frequency shift for the idler photons, with $\Delta\Omega$ large enough to rule out single-photon interference. The frequency-shifted and the frequency-unshifted paths interfere at a 50/50 beam splitter and acquire a phase difference of $\phi_S$ ($\phi_I$) within the signal (idler) interferometer. The outputs from both MZIs are sent to dispersive elements that impose second-order dispersions with equal magnitudes but opposite signs. The dispersed signal and idler photons are then detected by superconducting nanowire single-photon detectors (SNSPDs) and their timing coincidences are recorded. The second-order dispersions imposed by the dispersive elements correlates the frequency content of the inputs to their measured arrival times, thus effectively converting the performed time-domain measurement result to a frequency-domain measurement. The opposite signs of the two dispersive elements, together with nonlocal dispersion cancellation, recover the signal-idler frequency coincidences as signal-idler timing coincidences and thus distinguish between different signal-idler sum frequencies. 

The biphoton for time-energy entangled photon pairs produced by cw pumped SPDC can be written in its time-domain representation as \cite{SI}:
\begin{equation} \label{eq1}
|\psi\rangle_{SI} \propto \int\!{\rm d}t_-\,\psi_{SI}(t_-)|t_++t_-/2\rangle_S|t_+-t_-/2\rangle_I,
\end{equation}
where $t_+ = (t_S+t_I)/2$ and $t_- = (t_S-t_I)$, with $t_S$  $(t_I)$ representing the time for the signal (idler) photon. $\psi_{SI}(t_-)$ is the joint temporal amplitude and its magnitude squared is the joint temporal intensity ${\rm JTI}(t_-) = |\psi_{SI}(t_-)|^2$. The CFI's coincidence probability is given by \cite{SI}
\begin{equation} \label{eq2}
P_{\rm CFI}(\phi_T) = \frac{\eta^2}{8}\left(1+ \int\!{\rm d}t_-\,{\rm JTI}(t_-)\cos(\Delta\Omega t_- + \phi_T)\right)\,,
\end{equation}
where $\phi_T = \phi_S + \phi_I$ is the sum of the signal and idler MZI phase differences in the CFI and $\eta$ is the measurement efficiency in each MZI. The resulting visibility is
\begin{equation} \label{eq3}
V_{\rm CFI} = \int\!{\rm d}t_-\,{\rm JTI}(t_-)\cos(\Delta\Omega t_-).
\end{equation}
This visibility result is similar to those obtained in Franson interferometry for time-energy entangled photons \cite{franson1989bell} and the CHSH test with polarization-entangled photons \cite{clauser1969proposed} in that   the CFI is in the same class of quantum measurements for testing the violation of local hidden-variable theory and quantifying the nonlocal feature of entanglement.

\begin{figure*}
    \centering
    \includegraphics[scale=0.85]{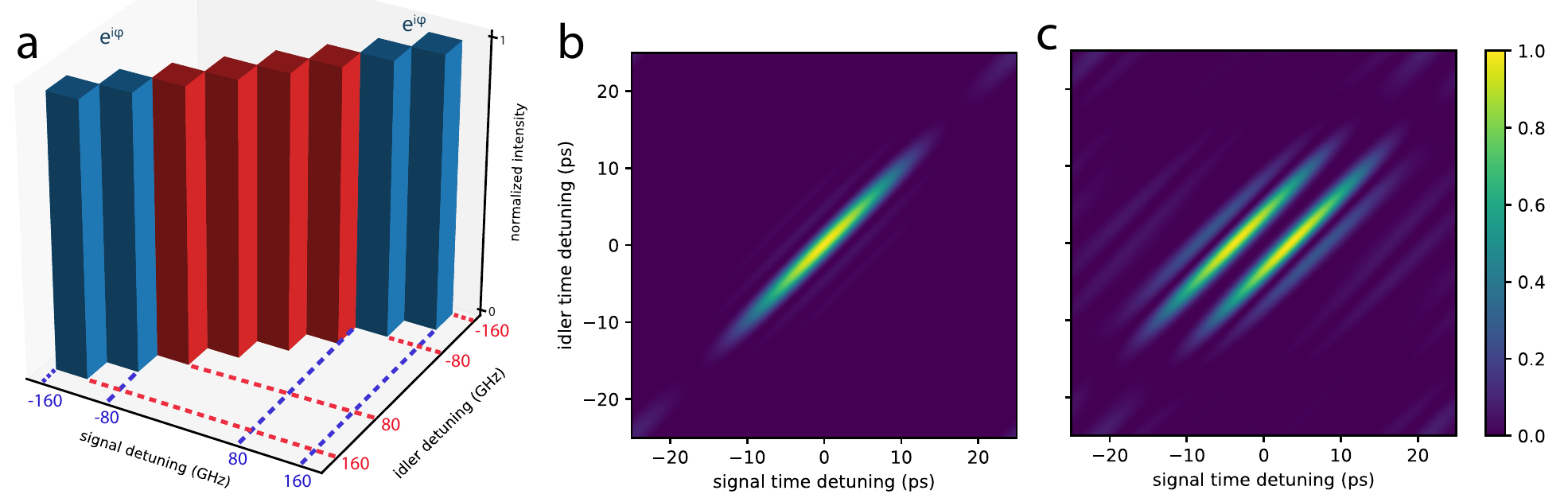}
    \caption{\textbf{a}, JSI calculated for a biphoton with 320-GHz-wide flat-top spectrum. Spectral phase $\phi$ (none, set to 0) applied to blue (red) shaded region outside (within) the $\pm 80$\,GHz span of signal and idler frequency detuning, showing no $\phi$ dependence.  \textbf{b}, JTI of same biphoton state with $\phi$ = 0 or $2\pi$. c. JTI of same biphoton state with $\phi$ = $\pi$. Maximum of JTI normalized to 1.}
    \label{fig3}
\end{figure*}

To demonstrate conjugate-Franson interferometry, we built a CFI as shown in the experimental schematic of Fig.~\ref{fig1}a with inputs of time-energy entangled photon pairs generated by cw pumped SPDC from a type-II phase-matched periodically-poled potassium titanyl phosphate (PPKTP) waveguide. The signal and idler photons were separated using a fiber polarizing beam splitter and sent to their respective MZIs. We repurposed two quadrature phase-shift keying (QPSK) modulators as the frequency shifters and set the frequency shift at $\pm \Delta \Omega/2\pi=\pm 15.65$\,GHz \cite{chen2021single}. We first characterized the frequency-shifted outputs from both frequency shifters using a narrowband cw laser at 1560 nm, as shown in Fig.~\ref{fig1}b. Within the desired frequency range from $-\Delta \Omega$ to $\Delta \Omega$, a minimum of 25 dB carrier-to-sideband ratio was achieved for both blue and red frequency shifters. The outputs from the signal and idler MZIs were sent to fiber Bragg-grating dispersion modules that imposed equal magnitude but opposite sign dispersions of $\pm$10\,ns/nm, after which the photons were detected using SNSPDs and the detections were time-tagged electronically. 

Because the SPDC signal-idler photon pairs are time-energy entangled, the imposed opposite dispersions cancel and their arrival times remain correlated \cite{lee2014entanglement}. Nevertheless, the existence of dispersion reveals the incoming photons' frequency information. The resolution of our frequency-domain measurement is 1.8\,GHz, which is determined by the detectors' timing jitter and the amount of applied dispersion. A sample signal-idler coincidence measurement from the CFI is shown in Fig.~\ref{fig1}c. The locations of the coincidence peaks correspond to the signal-idler sum frequencies which in turn indicate the possible paths the signal and idler photon have traveled. There are four possible path configurations as signal and idler photons can travel along either the frequency-shifted or the frequency-unshifted arms. The two side peaks correspond to the case in which only one of the signal and idler photons has been frequency shifted such that the signal-idler frequency sum is detuned by $\pm\Delta\Omega/2\pi=\pm 15.65$\,GHz. For the center peak the sum frequency remains unchanged, requiring that both photons travel along their frequency-unshifted arms or they both go through their respective frequency shifters. The two different paths are indistinguishable and they interfere as a function of the MZI phase sum $\phi_T$, producing the CFI's nonlocal coincidence interference similar to that of the Franson interferometer. We note that if the dispersion modules were not present, the three peaks could not be separated and the maximum interference visibility achievable would be limited to 50\%.  

We observed that the center coincidence peak of Fig.~\ref{fig1}c varied as a function of the phase sum $\phi_T$. The CFI was thermally insulated but we still observed that the center coincidence peak changed its magnitude due to residual thermal drift at an estimated rate of 0.3\,rad/min for $\phi_T$. We recorded the signal-idler coincidences and plotted the coincidence counts of the center peak as a function of the accumulated phase sum $\phi_T$, as shown in Fig.~\ref{fig2}a. The result shows a clear oscillatory signature as a function of the phase drift. To eliminate the possibility that the change of the coincidence counts was caused by changes of the photon flux, we also recorded the singles rates of both detectors at the same time during the coincidence measurement, as shown in Fig.~\ref{fig2}b. The measured singles rates remain constant throughout the thermal drift duration and show that the oscillatory fringe is not a result of single-photon interference. 

To obtain an accurate value for the CFI's interference visibility, we attached a piezoelectric transducer (PZT) stack to the signal MZI's frequency-unshifted arm as a fiber stretcher to impose a controllable phase shift on $\phi_S$. We repeatedly scanned $\phi_S$ from 0 to $2\pi$ while keeping $\phi_I$ constant. The fringe visibility was calculated based on the observed minimum and maximum coincidence counts within each phase scan. We obtain a CFI visibility of $96\pm1\%$ based on 23 phase-scan measurements and an uncertainty of 1 standard deviation. We estimate that degradation of our CFI visibility measurements was due to phase fluctuations of the CFI (1.2\%), modulators' extra sidebands ($0.7\%$), modulator dispersion ($0.5\%$), dark counts and noise background ($0.5\%$), and SPDC multi-pair events ($0.4\%$). The achieved visibility validates the quantum nonlocal correlation between our SPDC photon pairs, surpassing the quantum-classical threshold of $1/\sqrt{2}=$70.7\% by $\sim$25 standard deviations.  Although our current measurement setup is affected by the post-selection loophole, it can be modified to match the two side peaks temporally and eliminate the post-selection loophole \cite{vedovato2018postselection}. Furthermore, our 96\% measured visibility is on par with the 96\% visibility threshold set by a modified inequality that patches the post-selection loophole \cite{aerts1999two}. As a result, the high CFI visibility confirms that our photon-pair source indeed produces time-energy entanglement and validates conjugate-Franson interferometry's being a promising quantum measurement technique for certifying time-energy entanglement.

To show that the CFI brings new capability to the increasingly expanding photonic quantum toolbox, we demonstrate that the CFI visibility is sensitive to the spectral phase of a biphoton state, something which cannot be sensed by Franson or Hong-Ou-Mandel interferometers. First consider a cw pumped SPDC source generating a time-energy entangled biphoton state with a flat spectrum spanning 320\,GHz and no spectral phase variation, i.e., its frequency-domain description is
\begin{equation}
|\psi^{(1)}\rangle_{SI} \mbox{$\propto$} \int_{-\omega_{\rm max}}^{\omega_{\rm max}}\!{\rm d}\omega\,\Psi^{(1)}_{SI}(\omega)|\omega_{S_0}+\omega\rangle_S |\omega_{I_0}-\omega\rangle_I\,,
\end{equation}
where $\Psi^{(1)}_{SI}(\omega) = 1/\sqrt{2\omega_{\rm max}}$ is its joint spectral amplitude (JSA), $\omega_{S_0}$ ($\omega_{I_0}$) is the signal (idler) center frequency, and $\omega$ is the state's frequency detuning with a range of $\pm\omega_{\rm max}$ where $\omega_{\rm max}/2\pi = 160$\,GHz. Now consider the state $|\psi^{(2)}\rangle_{SI}$ whose JSA is
\begin{equation}
\Psi^{(2)}_{SI}(\omega) = \left\{\begin{array}{ll}
1/\sqrt{2\omega_{\rm max}}, & \mbox{for $|\omega| \le \omega_1$}\\[.05in]
 e^{i\phi}/\sqrt{2\omega_{\rm max}}, & \mbox{ for $\omega_1 < |\omega| \le \omega_{\rm max}$},
 \end{array}\right.
\end{equation} 
where $\omega_1/2\pi = 80$\,GHz. 

\begin{figure}[t!]
    \centering
    \includegraphics[scale=0.62]{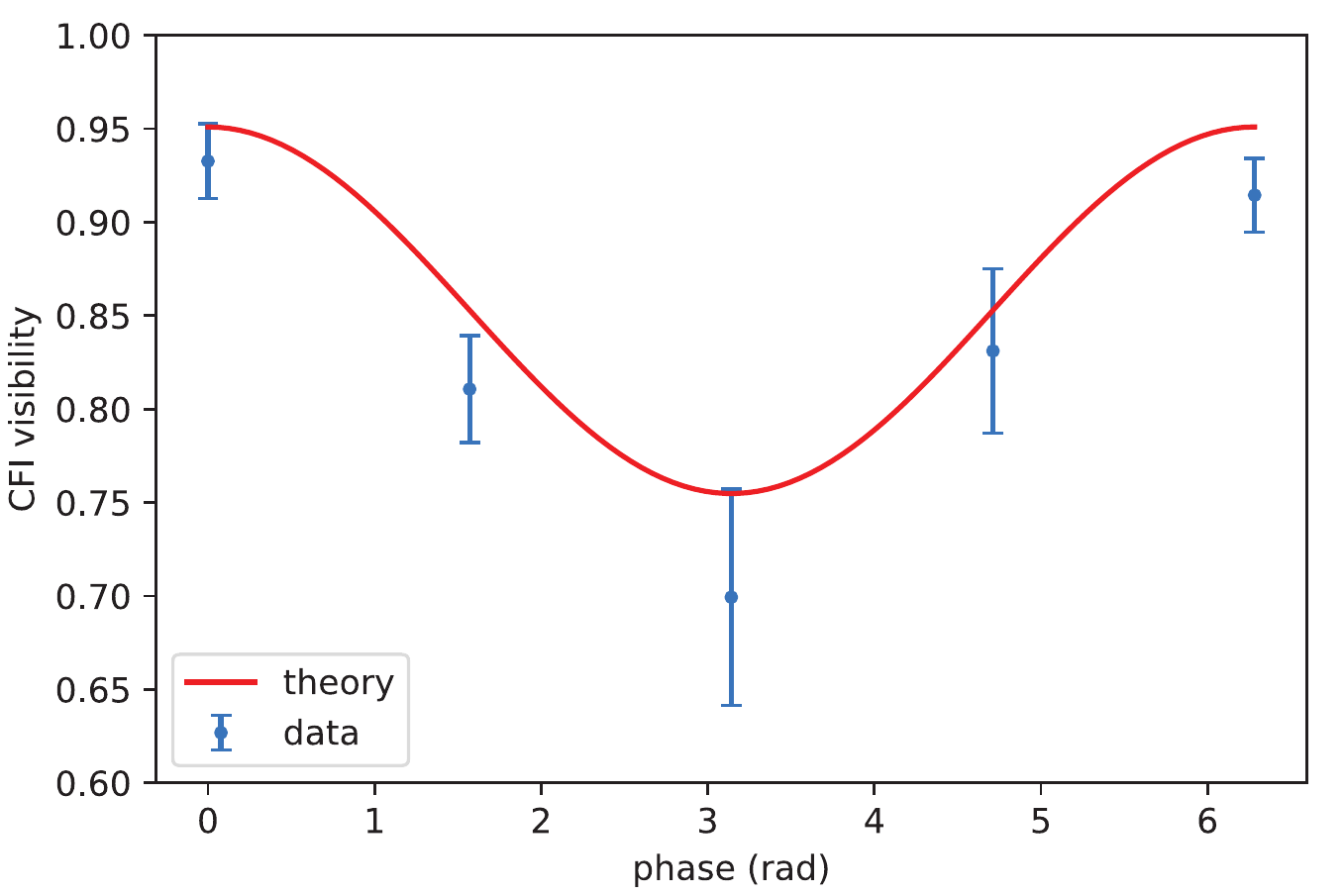}
    \caption{Conjugate-Franson fringe visibility as a function of applied spectral phase $\phi$ of equation 5. Measured data points (blue) follow closely the calculated values (red) obtained from equation 3 with a rectangular spectrum of 320\,GHz span shown in Fig.~\ref{fig3}a.}
    \label{fig4}
\end{figure}

Although $|\psi^{(2)}\rangle_{SI}$ differs from $|\psi^{(1)}\rangle_{SI}$ when $0 < \phi < 2\pi$, these states cannot be distinguished by Franson or Hong-Ou-Mandel interference because $|\psi^{(2)}\rangle_{SI}$ and $|\psi^{(1)}\rangle_{SI}$ have identical JSIs, as shown in Fig.~3a, and those interferometers' interference patterns are determined by the JSI. On the other hand, the JTIs of $|\psi^{(2)}\rangle_{SI}$ and $|\psi^{(1)}\rangle_{SI}$ are different, because of JTI's spectral phase dependence. This difference is shown in Fig.~\ref{fig3}b and c, which display the JTIs of $|\psi^{(2)}\rangle$ for $\phi=0$ and $\pi$, respectively., with the former also being the JTI of $|\psi^{(1)}\rangle_{SI}.$ Equation~\ref{eq3} indicates that the CFI visibility is a function of the JTI and thus sensitive to spectral phase. Our theoretical calculation for the CFI visibility yields 95.1\% for $|\psi^{(1)}\rangle_{SI}$ and 75.5\% for $|\psi^{(2)}\rangle_{SI}$ with $\phi = \pi$. This represents a $\sim$20\% drop in CFI visibility that should be measurable experimentally.

We used a type-0 phase-matched periodically poled lithium niobate (PPLN) crystal to generate time-energy entangled photon pairs with a flat spectrum across the telecommunication C band. We applied a programmable amplitude and phase spectral filter to shape the signal and idler spectra to be rectangular with a 320\,GHz bandwidth and to impose an adjustable phase $e^{i\phi}$ on both signal and idler light for frequency detuning $|\omega|/2\pi$ between 80 to 160\,GHz, thus producing the biphoton state $|\psi^{(2)}\rangle_{SI}$. We measured the CFI visibility at $\phi=0$, $\pi/2$, $\pi$, $3\pi/2$, and $2\pi$ and Fig.~\ref{fig4} displays our results along with the theoretically calculated values. Because $\phi =0$ or $2\pi$ makes $|\psi^{(2)}\rangle_{SI} = |\psi^{(1)}\rangle_{SI}$, the $93.2 \pm 2.0\%$ visibility we obtained for $\phi =0$ and the $91.4 \pm 2.0\%$ we got for $\phi = 2\pi$, with the uncertainty value being the standard deviation of 3 measurements, are consistent with that equivalence. Figure~\ref{fig4} shows that the CFI visibility degrades when spectral phase variation was introduced, reaching a minimum visibility of $72.0 \pm 3.1\%$ for $\phi=\pi$, in good agreement with our calculation. In this simple example, the substantial visibility reduction of 21.2\% from $\phi=0$ to $\phi=\pi$ clearly confirms the ability of the CFI to distinguish between states with different spectral phase content. Our experimental results show that conjugate-Franson interferometry can be used not only for quantifying time-energy entanglement of biphotons but also for detecting their spectral phase differences, which is helpful in characterizing entangled systems with high-dimensional encoding \cite{lingaraju2019quantum, chang2021648}.

In summary, we reported experimental realization of the conjugate-Franson interferometer, demonstrating a CFI visibility of $96\pm1$\% without any background subtraction for time-energy entangled photon pairs generated by cw pumped SPDC. The achieved visibility surpasses the quantum-classical threshold of $\sim$71\% by 25 standard deviations and clearly validates the quantum entanglement feature between the SPDC signal and idler photons. To illustrate its application potential, we utilized the CFI as an enabling quantum measurement technique to distinguish two biphoton states with identical joint spectral intensities but different joint temporal intensities due to spectral phase variation. By introducing an adjustable spectral phase shift to a cw pumped SPDC biphoton state, we observed a significant CFI visibility drop of 21\% between the two biphoton states, matching our theoretical calculations. Our results show that conjugate-Franson interferometry quantifies correlation in the time domain and is complementary to the well-known Franson interferometry. The CFI's dependence on the joint temporal intensity makes it a valuable addition to the suite of quantum measurement techniques for entanglement characterization and verification. 


\section{Method}
The PPKTP waveguide was type-II phase-matched and pumped by a 780\,nm cw laser. The orthogonally-polarized signal and idler photons were nondegenerate with $\sim$200\,GHz offset between their center frequencies and each had a full-width at half-maximum (FWHM) bandwidth of 320\,GHz. We used polarization control paddles and polarizing beam splitters to balance the flux between the frequency-shifted arm and the frequency-unshifted arm within each MZI of the CFI. The frequency shifters were dual-drive quadrature phase-shift keying (QPSK) modulators (Fujitsu FTM7961EX) operating in a configuration for single sideband generation \cite{chen2021single}. The radio frequency (RF) electrical inputs to both modulators were derived from the same RF synthesizer. For each modulator, the 15.65\,GHz RF signal was amplified and split by a $0^\circ$-phase 50/50 power splitter to serve as inputs to the frequency shifters. The cables connecting the two power-splitter outputs to the modulator had a 10.96\,cm length difference so that the split RF signals had a $\pi/2$ phase shift when they arrived at the modulator to satisfy the single-sideband generation condition. The signal's frequency shifter was configured to blue shift its input while the idler's shifter was configured to red shift its input. Tunable air gaps on translation stages were used to match the path lengths between the frequency-shifted and frequency-unshifted arms. We used a 50-nm-bandwidth superluminescent diode (SLD) at 1560\,nm to ensure the two path lengths were well matched. The path-length mismatch was upper bounded by the SLD source's 16\,$\mu$m coherence length, which is much less than the expected $\sim 200\,\mu$m biphoton coherence length. 

The polarization between the two arms of each MZI was calibrated to be the same using classical light to ensure optimal interference at the 50/50 beam splitter. During operation, the signal's MZI had an 18.6\,dB insertion loss and the idler's MZI had a 22.7\,dB insertion loss. These high insertion losses were mainly due to the low conversion efficiencies of the frequency shifters~\cite{chen2021single}. The different insertion losses of the two MZIs was caused by performance difference of the two frequency shifters and tunable air gaps. The two fiber Bragg-grating dispersion modules had 3\,dB insertion loss and passband from 1557.37\,nm (192.50\,THz) to 1562.74\,nm (191.84\,THz). After the dispersion modules, the photons were detected with WSi SNSPDs with $\sim$80\% system efficiency and 120\,ps timing jitter. The detected signal and idler spectral ranges were limited by the dispersion modules' 660\,GHz passband. The detection events were time-tagged using a time-tagger (Hydraharp 400) with 128\,ps timing resolution.

We placed the CFI in a custom-built two-stage thermal enclosure. Both the outer and inner layers were made from cardboard and thermal-isolation foam. This passive thermal enclosure slowed down the ambient thermal fluctuation and also restricted the inside air current flow so that the phase of the fiber interferometer was relatively stable for the duration of measurements. During measurement, the temperature outside the enclosure was kept reasonably stable. Nevertheless, we observed that residual environmental fluctuations imposed a phase drift on the CFI's $\phi_T$ at a $\sim$0.3\,rad/min rate, which we measured by monitoring the power drift of both MZIs simultaneously. The observed phase drift was relatively small so that the phase during measurement could be approximated by a constant. The signal-idler coincidences were recorded as the CFI underwent the thermal phase drift process and each coincidence data point was integrated for 30 seconds. 

For measuring the CFI visibility, we focused on obtaining the coincidence counts near the phase locations where the maximum and minimum coincidence rates would occur. We used a 150\,V PZT stack to serve as a controllable phase shifter by stretching the fiber length on the frequency-unshifted arm of the signal MZI. The stretched fiber increased the optical path length and introduced more precise phase shifts. Using the PZT stack, we were able to apply phase shifts from 0 to $2\pi$ when we applied voltages from 0 to 120\,V. We changed the phase from 0 to 2$\pi$ adaptively to measure the coincidences near their maximum and minimum values. When the measured coincidence was within 10\% of either the maximum or minimum, we applied phase-change steps of 0.15\,rad. When the measured coincidence was outside of the 10\% range of the target maximum or minimum counts, a larger phase-change step of 0.52 rad was used. This adaptive strategy allowed us to capture the maximum and minimum coincidences in a more efficient and controllable manner.

For the measurement to distinguish between two biphoton states with different spectral phases, we used a PPLN crystal that was type-0 phase-matched and pumped by a 780\,nm cw laser. A 50/50 beam splitter was used to separate the co-polarized signal and idler photons that incurred a 3\,dB loss for postselected signal-idler coincidence measurements. The signal and idler had flat spectra across the telecommunication C band. We used a Finisar waveshaper 1000S as the programmable spectral filter to control both the amplitude and phase of signal-idler joint spectral amplitude.

\begin{acknowledgments}
This work was supported by the National Science Foundation under grant number 1741707. The authors thank Catherine Lee and Dirk Englund for providing the Proximion dispersion compensation modules, and Boya Ye for generating the three-dimensional experimental schematic.
\end{acknowledgments}

\end{document}


\title{Supplemental Material for Experimental Demonstration of Conjugate-Franson Interferometry}
\author{Changchen Chen, Jeffrey H. Shapiro, and Franco N.C. Wong}
\affiliation{Research Laboratory of Electronics, Massachusetts Institute of Technology, Cambridge, MA 02139}
\date{\today}

\maketitle

\section{Introduction}
The conjugate Franson interferometer (CFI) was introduced in Ref.~\cite{Zhang2014} as a tool for securing high-dimensional quantum key distribution based on time-energy entangled biphotons.  That paper's Supplemental Material includes a detailed derivation of the CFI's coincidence-count behavior.  Thus we can content ourselves with a briefer presentation that gets at an essential feature of the CFI that was not made explicit in Ref.~\cite{Zhang2014}, i.e., the CFI's coincidence behavior is controlled by the biphoton state's joint temporal intensity (JTI).  As such, the CFI complements the conventional Franson interferometer (FI), whose coincidence behavior is controlled by the biphoton state's joint spectral intensity (JSI).  The biphoton's JSI is the squared magnitude of its joint spectral amplitude (JSA), which is the biphoton's properly normalized frequency-domain wave function.  Similarly, the biphoton's JTI is the squared magnitude of its joint temporal amplitude (JTA), which is the biphoton's properly normalized time-domain wave function.  It follows that knowing both the JSI and the JTI will allow the biphoton's full state to be determined by applying standard phase-retrieval techniques to recover the JSA's missing spectral phase, see, e.g.,~\cite{GSalgorithm}.   

\section{Preliminaries}
We are interested in single spatial mode signal and idler fields produced by a type-II or type-0 phase matched spontaneous parametric downconverter.  The scalar, photon-units, positive-frequency field operators for the relevant polarizations of the signal and idler will be taken to be
$\hat{E}_S^{(+)}(t)$ and $\hat{E}_I^{(+)}(t)$ with the usual $\delta$-function commutators,
\begin{equation}
[\hat{E}_K^{(+)}(t),\hat{E}_K^{(+)\dagger}(u)] = \delta(t-u), \mbox{ for $K = S,I.$}
\end{equation}
For what will follow it will be valuable to have these operators' frequency-domain decompositions~\cite{footnote1, footnote2},
\begin{equation}
\hat{E}_S^{(+)}(t) = \int\!\frac{{\rm d}\omega_S}{2\pi}\,\mathcal{E}_S(\omega_S)e^{-i(\omega_{S_0} + \omega_S)t}\mbox{\,\, and\,\, }
\hat{E}_I^{(+)}(t) = \int\!\frac{{\rm d}\omega_I}{2\pi}\,\mathcal{E}_I(\omega_I)e^{-i(\omega_{I_0} - \omega_I)t},
\end{equation}
where $\omega_{S_0}$ ($\omega_{I_0}$) is the center frequency of the signal (idler).

Our interest will be in biphoton states of these signal and idler, i.e., states of the form
\begin{equation}
|\psi\rangle_{SI} = \frac{1}{2\pi}\int\!{\rm d}\omega_S \int\!{\rm d}\omega_I\,\Psi_{SI}(\omega_S,\omega_I)
|\omega_{S_0}+\omega_S\rangle_S|\omega_{I_0}-\omega_I\rangle_I\, . \\[.05in]
\label{freqRep}
\end{equation}
Here, $|\omega_{S_0}+\omega_S\rangle_S$ and $|\omega_{I_0}-\omega_I\rangle_I$ are signal and idler states consisting of  single photons at detunings $\omega_S$ and $-\omega_I$, respectively, from those field's center frequencies.  The preceding states are properly normalized, viz., ${}_{SI}\langle \psi|\psi\rangle_{SI} = 1$.  Hence, taking
\begin{equation}
{}_S\langle \omega_{S_0} + \omega_S|\omega_{S_0}+\omega'_S\rangle_S  = 2\pi\,\delta(\omega_S-\omega'_S) \mbox{\,\, and\,\, }
{}_I\langle \omega_{I_0} - \omega_I|\omega_{I_0}-\omega'_I\rangle_I  = 2\pi\,\delta(\omega_I-\omega'_I),
\label{wwprime}
\end{equation}
we find that $\int\!{\rm d}\omega_S\int\!{\rm d}\omega_I\, |\Psi_{SI}(\omega_S,\omega_I)|^2 = 1$.  In other words, this biphoton's JSA is
\begin{equation}
{\rm JSA}(\omega_S,\omega_I) = \Psi_{SI}(\omega_S,\omega_I),
\end{equation}
and its JSI is
\begin{equation}
{\rm JSI}(\omega_S,\omega_I) = |\Psi_{SI}(\omega_S,\omega_I)|^2.
\end{equation}

For a time-domain representation of this biphoton, we introduce signal and idler states $|t_S\rangle_S$ and $|t_I\rangle_I$ consisting of single photons at times $t_S$ and $t_I$, respectively.  These states are the Fourier duals of $|\omega_{S_0}+\omega_S\rangle_S$ and $|\omega_{I_0}-\omega_I\rangle_I$, i.e., we have
\begin{equation}
|t_S\rangle_S = \int\!\frac{{\rm d}\omega_S}{2\pi}\,e^{i(\omega_{S_0}+\omega_S)t_S}|\omega_{S_0}+\omega_S\rangle_S \mbox{\,\, and\,\, } 
|t_I\rangle_I = \int\!\frac{{\rm d}\omega_I}{2\pi}\,e^{i(\omega_{I_0}-\omega_I)t_I}|\omega_{I_0}-\omega_I\rangle_I,
\label{tfromw}
\end{equation}
and
\begin{equation}
|\omega_S\rangle_S = \int\!{\rm d}t_S\,e^{-i(\omega_{S_0}+\omega_S)t_S}|t_S\rangle_S \mbox{\,\, and\,\, }
|\omega_I\rangle_I = \int\!{\rm d}t_I\,e^{-i(\omega_{I_0}-\omega_I)t_I}|t_I\rangle_I,
\label{wfromt}
\end{equation}
from which we get
\begin{equation}
{}_K\langle t_K|t'_K\rangle_K = \delta(t_K-t'_K), \mbox{ for $K = S,I$.}
\label{ttprime}
\end{equation}
Then, direct evaluation using Eqs.~(\ref{freqRep}), (\ref{wwprime}), and (\ref{tfromw}) gives
\begin{equation}
{}_S\langle t_S|{}_I\langle t_I|\psi\rangle_{SI} = \frac{e^{-i(\omega_{S_0}t_S+\omega_{I_0}t_I)}}{2\pi}
\int\!{\rm d}\omega_S\int\!{\rm d}\omega_I\,\Psi_{SI}(\omega_S,\omega_I)e^{-i(\omega_St_S - \omega_It_I)},
\end{equation}
which implies that
\begin{equation}
|\psi\rangle_{SI} =  \int\!{\rm d}t_S \int\!{\rm d}t_I\,\psi_{SI}(t_S,t_I)
|t_S\rangle_S|t_I\rangle_I,
\end{equation}
with
\begin{equation}
\psi_{SI}(t_S,t_I) = \frac{e^{-i(\omega_{S_0}t_S + \omega_{I_0}t_I)}}{2\pi}
\int\!{\rm d}\omega_S\int\!{\rm d}\omega_I\,\Psi_{SI}(\omega_S,\omega_I)e^{-i(\omega_St_S - \omega_It_I)}
\label{psifromPsi}
\end{equation}
satisfying $\int\!{\rm d}t_S\int\!{\rm d}t_I |\psi_{SI}(t_S,t_I)|^2 = 1$.  It is now easily seen that the  biphoton's JTA and JTI are 
\begin{equation}
{\rm JTA}(t_S,t_I) = \psi_{SI}(t_S,t_I),
\end{equation}
and
\begin{equation}
{\rm JTI}(t_S,t_I) = |\psi_{SI}(t_S,t_I)|^2.
\end{equation}

Later we shall employ the Gaussian biphoton wave functions,
\begin{equation}
\Psi_{SI}(\omega_S,\omega_I) = \frac{e^{-(\omega_S-\omega_I)^2\sigma_{\rm coh}^2}\,e^{-(\omega_S+\omega_I)^2\sigma_{\rm cor}^2/4}}{\sqrt{\pi/2\sigma_{\rm coh}\sigma_{\rm cor}}},
\end{equation}
and
\begin{equation}
\psi_{SI}(t_S,t_I) = \frac{e^{-i(\omega_{S_0}t_S+\omega_{I_0}t_I)}\,e^{-(t_S+t_I)^2/16\sigma^2_{\rm coh}}\,e^{-(t_S-t_I)^2/4\sigma^2_{\rm cor}}}{\sqrt{2\pi \sigma_{\rm coh}\sigma_{\rm cor}}},
\label{Gauss_psi}
\end{equation}
where the root-mean-square (rms) coherence time, $\sigma_{\rm coh}$, is determined by the downconverter's pump linewidth and the rms correlation time, $\sigma_{\rm cor}$, is determined by the downconverter's phase-matching bandwidth.  This biphoton can be realized
by engineered phase matching of a periodically-poled nonlinear crystal, see, e.g.,~\cite{Chen2019}.

\section{Coincidence Probability of Conjugate Franson Interferometry}
To instantiate the CFI configuration shown in Fig. 1 of the main text, let the downconverter's signal beam undergo the $\Delta \Omega >0$ frequency shift and normal dispersion, while the downconverter's idler beam undergoes the $-\Delta\Omega < 0$ frequency shift and anomalous dispersion.  Taking all the optics to be lossless, we then have that the positive-frequency field operators illuminating the single-photon detectors in Fig. 1 are
\begin{equation}
\hat{E}_{S'}^{(+)}(t) = \frac{1}{2}\int\!\frac{{\rm d}\omega_S}{2\pi}\,[\mathcal{E}_S(\omega_S) + \mathcal{E}_S(\omega_S+\Delta\Omega)e^{i\phi_S}]e^{i\beta_2\omega_S^2/2}e^{-i(\omega_{S_0}+\omega_S)t}
\end{equation}
and
\begin{equation}
\hat{E}_{I'}^{(+)}(t) = \frac{1}{2}\int\!\frac{{\rm d}\omega_I}{2\pi}\,[\mathcal{E}_I(\omega_I) + \mathcal{E}_I(\omega_I+\Delta\Omega)e^{i\phi_I}]e^{-i\beta_2\omega_I^2/2}e^{-i(\omega_{I_0}-\omega_I)t}.
\end{equation}
As shown in Ref.~\cite{Zhang2014}, for sufficiently high frequency shifts and a sufficiently high dispersion coefficient, $\beta_2$, there will not be any second-order interference and, in the absence of dark counts, the probability of registering a coincidence from a biphoton emitted by the downconverter is 
\begin{equation}
P_{\rm CFI}(\phi_S,\phi_I) = \frac{\eta^2}{8}\left(1+\int\!{\rm d}\omega_S\int\!{\rm d}\omega_I\,{\rm Re}[\Psi_{SI}^*(\omega_S,\omega_I)\Psi_{SI}(\omega_S+\Delta\Omega,\omega_I+\Delta\Omega)e^{i(\phi_S+\phi_I)}]\right),
\label{coincprob1}
\end{equation}
where $\eta$ is the detectors' quantum efficiency.  

To rewrite Eq.~(\ref{coincprob1}) in terms of $\psi_{SI}(t_S,t_I)$, we first invert Eq.~(\ref{psifromPsi}) to obtain
\begin{equation}
\Psi_{SI}(\omega_S,\omega_I) = \frac{1}{2\pi} \int\!{\rm d}t_S\int\!{\rm d}t_I\,\psi_{SI}(t_S,t_I)e^{i[(\omega_{S_0}+\omega_S)t_S + (\omega_{I_0}-\omega_I)t_I]}.
\end{equation}
Using this result in Eq.~(\ref{coincprob1}) gives us the result we are seeking,
\begin{align}
P_{\rm CFI}(\phi_S,\phi_I) &= \frac{\eta^2}{8}\left(1 + \int\!{\rm d}t_S\int\!{\rm d}t_I\,
|\psi_{SI}(t_S,t_I)|^2\cos[\Delta\Omega(t_S-t_I) + (\phi_S +\phi_I)]\right) \\[.05in]
&= \frac{\eta^2}{8}\left(1 + \int\!{\rm d}t_S\int\!{\rm d}t_I\,
{\rm JTI}(t_S,t_I)\cos[\Delta\Omega(t_S-t_I) + \phi_T]\right),
\label{coincprob2}
\end{align} 
where $\phi_T \equiv \phi_S + \phi_I$ is the interferometer's phase sum. As an illustration of the coincidence probability's behavior, let us evaluate Eq.~(\ref{coincprob2}) using $\psi(t_S,t_I)$ from Eq.~(\ref{Gauss_psi}).  The double integral is easily performed if we change to sum and difference coordinates, i.e., $t_+ \equiv (t_S+t_I)/2$ and $t_- \equiv t_S-t_I$.  The result we obtain is
\begin{equation}
P_{\rm CFI}(\phi_T) = \frac{\eta^2[1+e^{-\Delta\Omega \sigma^2_{\rm cor}/2}\cos(\phi_T)]}{8},
\end{equation}
which implies an interference fringe visibility  
\begin{equation}
V_{\rm CFI} \equiv \frac{\max_{\phi_T}[P_{\rm CFI}(\phi_T)] - \min_{\phi_T}[P_{\rm CFI}(\phi_T)]}{\max_{\phi_T}[P_{\rm CFI}(\phi_T)] + \min_{\phi_T}[P_{\rm CFI}(\phi_T)]} = 
e^{-\Delta\Omega^2\sigma^2_{\rm cor}/2} \approx 1, \mbox{ for $\Delta\Omega \ll 1/\sigma_{\rm cor}$.}
\end{equation}

The preceding analysis must be modified to treat the case of ideal continuous-wave (cw) pumped downconversion.  The biphoton such an arrangement generates is a state
\begin{equation}
|\psi\rangle_{SI} \propto \int\!{\rm d}\omega\,\Psi_{SI}(\omega)|\omega_{S_0}+\omega\rangle_S |\omega_{I_0}-\omega\rangle_I
\end{equation}
that cannot be normalized, i.e., ${}_{SI}\langle \psi|\psi\rangle_{SI} = \infty$.  Nevertheless, we can normalize $\Psi_{SI}(\omega)$ to serve as this biphoton's JSA for $\omega_S = -\omega_I = \omega$ and $|\Psi_{SI}(\omega)|^2$ as its JSI for $\omega_S = -\omega_I = \omega$.  Furthermore, by neglecting the unimportant---insofar as $P_{\rm CFI}(\phi_T)$ is concerned---phase factor $e^{-i(\omega_{S_0}t_S + \omega_{I_0}t_I)}$, we can define
\begin{equation}
\psi_{SI}(t_-) = \frac{1}{\sqrt{4\pi}}\int\!{\rm d}\omega\,\Psi_{SI}(\omega)e^{-i\omega t_-/2},
\end{equation}
so that the unnormalizable $|\psi\rangle_{SI}$ can be rewritten as
\begin{equation}
|\psi\rangle_{SI} \propto \int\!{\rm d}t_-\,\psi_{SI}(t_-)|t_++t_-/2\rangle_S|t_+-t_-/2\rangle_I,
\end{equation}
where, as before, $t_+ = (t_S+t_I)/2$ and $t_- = (t_S-t_I)$.  The JTI for this cw case is then
${\rm JTI}(t_-) = |\psi_{SI}(t_-)|^2$ and the CFI's coincidence probability is
\begin{equation}
P_{\rm CFI}(\phi_T) = \frac{\eta^2}{8}\left(1+ \int\!{\rm d}t_-\,{\rm JTI}(t_-)\cos(\Delta\Omega t_- + \phi_T)\right).
\end{equation}
The resulting visibility is
\begin{equation}
V_{\rm CFI} = \int\!{\rm d}t_-\,{\rm JTI}(t_-)\cos(\Delta\Omega t_-).
\end{equation}

\section{Conjugate-Franson Interferometer Phase Stability Characterization}
\begin{figure}[h!]
    \centering
    \includegraphics[scale=0.45]{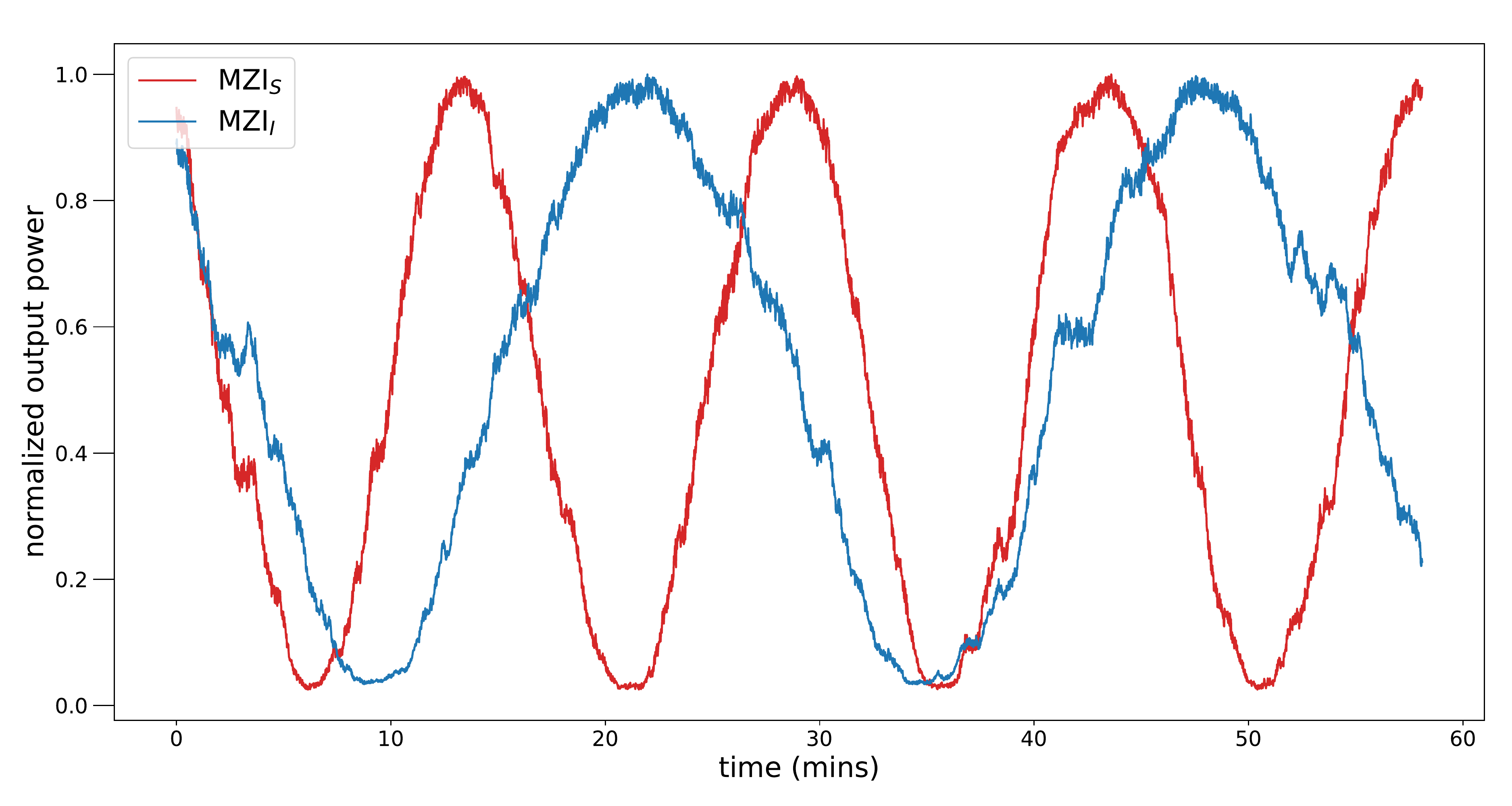}
    \caption{Normalized output power from MZIs measured at 1s interval. The red (blue) curve is the measured power for signal's (idler's) MZI.}
    \label{sfig_1}
\end{figure}

To characterize our CFI setup's phase stability, we simultaneously monitored the power variations in the signal and idler's balanced Mach-Zehnder interferometers (MZIs) with a 1560\,nm cw laser. In this phase stability measurement, frequency shift was not implemented. The measured results are shown in Fig. \ref{sfig_1}. From the MZIs' output powers, we calculated their relative phase changes $\Delta\phi_S$ and $\Delta\phi_I$. The change in the total phase shift is the sum of the two individual phase shift changes, i.e., $\Delta\phi_T = \Delta\phi_S + \Delta\phi_I$. Overall, the calculated average drift rate of the sum phase was $\Delta\phi_T/\Delta t$ is $\sim$0.3\,rad/min.

